\documentclass[aps, showpacs, prb,superscriptaddress,twocolumn,a4paper]{revtex4-1}
\usepackage{graphicx}
\usepackage{amsmath, amsfonts, amssymb, bm}
\usepackage{color}
\usepackage{multirow}
\usepackage[utf8]{inputenc}
\usepackage[english]{babel}
\usepackage{amsmath,array,amssymb,pst-all}
\usepackage{dsfont}
\usepackage{bbold}
\newcommand{\ket}[1]{\left| #1 \right\rangle}

\newcommand{\ketbra}[2]{\left| #1 \right\rangle\left\langle #2 \right|}

\newcommand{\w}{\omega}

\begin{document}

\title{Room Temperature Quantum Coherence vs. Electron Transfer in a Rhodanine Derivative Chromophore}
\author{Duvalier Madrid-Úsuga}
\altaffiliation{{\tt{duvalier.madrid@correounivalle.edu.co}}}
\address{Centre for Bioinformatics and Photonics (CIBioFi), Calle 13 No.~100-00, Edificio E20 No.~1069, Universidad del Valle, 760032 Cali, Colombia}
\address{Departamento de Física, Universidad del Valle, 760032 Cali, Colombia}
\author{Cristian E. Susa}
\altaffiliation{{\tt{cristiansusa@correo.unicordoba.edu.co}}}
\address{Centre for Bioinformatics and Photonics (CIBioFi), Calle 13 No.~100-00, Edificio E20 No.~1069, Universidad del Valle, 760032 Cali, Colombia}
\address{Departamento de Física y Electrónica, Universidad de Córdoba, 
230002 Montería, Colombia}
\author{John H. Reina}
\altaffiliation{{\tt{john.reina@correounivalle.edu.co}}}
\address{Centre for Bioinformatics and Photonics (CIBioFi), Calle 13 No.~100-00, Edificio E20 No.~1069, Universidad del Valle, 760032 Cali, Colombia}
\address{Departamento de Física, Universidad del Valle, 760032 Cali, Colombia}

  
\begin{abstract}
Understanding electron transfer in organic molecules is of great interest in quantum materials for light harvesting, energy conversion and integration of molecules into solar cells. This, however, poses  the challenge of designing specific optimal molecular structure for which the processes of ultrafast quantum coherence and electron transport are not so well understood. In this work, we investigate subpicosecond time scale quantum dynamics and electron transfer in an efficient electron acceptor Rhodanine chromophoric complex. We consider an open quantum system approach to model the complex-solvent interaction,  and compute the crossover from weak to strong dissipation on the reduced system dynamics for both a polar (Methanol) and a non polar solvent (Toluene). We show that the electron transfer rates are enhanced in the strong chromophore-solvent coupling regime, being the highest transfer rates those found at room temperature. Even though the computed dynamics are highly non-Markovian, and they
 may exhibit a quantum character up to hundreds of femtoseconds, we  show that quantum coherence does not necessarily optimise the electron transfer in the chromophore.

\end{abstract}
\maketitle

\section{Introduction}

Electron transfer (ET) mechanisms and dynamics are ubiquitous to fundamental processing of energy in biological and chemical systems that are essential to life. They are at the core of light harvesting for energy production in a variety of systems; e.g., biomolecular photosynthetic complexes, biosensors, and solar cells~\cite{engel2007,panitchayangkoon2010,collini2009, hwang2010,song2016, hedley2016} to name but a few. From a perspective of novel materials and technologies for solar energy conversion, organic photovoltaic (OPV) sytems have attracted significant attention~{\cite{sauve2015, pandey2010, ho2015}} due to the  enhancement of their power conversion efficiency (PCE)~{\cite{song2016, hedley2016}}, flexibility,manufacturing reduction costs, and high-temperature production of cells~{\cite{hughes2014,Insuasty2011385,madrid2018,cabrera2018, cabrera2017}}.
On a different facet, but complementary to this development, quantum coherence has, over the past decade, been experimentally linked to early-stage energy transfer in light harvesting photosynthetic complexes~\cite{engel2007,panitchayangkoon2010,collini2010,lee2007C, collini2009, hwang2010, tempelaar2015, roscioli2017,strumpfer2012e};  in particular, time-resolved ultrafast optical and electronic spectroscopies have  evidenced oscillations of exciton state populations lasting up to a few hundred femtoseconds, which have been attributed to quantum coherence emerging as a result of initially prepared laser pulses~\cite{engel2007,panitchayangkoon2010,collini2010,lee2007C, collini2009, hwang2010, tempelaar2015, roscioli2017}. This two-way breakthrough poses an intriguing question about the possible functional role of quantum coherent effects as a precursor of efficient ET in materials and biological systems under structural and energetic disorder at physiological conditions.

The above physical insight requires that ET reactions in such chemical or biological light harvesters must consider the effects due to electronic coupling, interactions with the environment, and intramolecular vibrational modes.In particular, photoinduced ultrafast energy and ET processes in OPV  devices require a quantum-mechanical treatment of such fundamental microscopic processes. Here, electron transfer occurs between electron-donor and electron-acceptor molecules immersed in a solvent, and such process can be modelled by considering a quantum system in interaction with a local environment (solvent), whose dynamics, control, and correlations are amenable to a description within  the framework of open quantum systems~{\cite{Weiss,breuer2002,RevModPhys,reina2014,reina2002,koch2016,eckel2009, thorwart2009,chen2009,reina2018,susa2014,melo2017,forgy2014,mazziotti2012,rebentrost2009}}. Donor-acceptor chromophoric complexes convey electronic coupling between sites on a scale that relies on the solvent reorganisation energy and defines different regimes for the complex coherent/incoherent dynamics. Due to the interplay between energy fluctuation and transport, here we account for the crossover from weak to strong dissipation and consider memory effects for the chromophore-solvent quantum dynamics~{\cite{yuan2009, farhat2017}}; these can be described by means of intensive computational techniques such as path-integral based formalisms~{\cite{makarov1994, makri1995, eckel2009,thorwart2009}}, Monte Carlo algorithms~{\cite{muhlbacher2003}}, Hierarchical Equations of Motion (HEOM)~{\cite{Ishizaki, tanimura2006, ishizaki2009, chen2015}}, and reaction-coordinate methods {\cite{bolhuis2000}}.

In this work, we consider an OPV light harvester (an efficient Rhodanine derivative chromophore~{\cite{Insuasty2011385,madrid2018}}),  and focus on theoretical aspects regarding its ultrafast carrier dynamics and the interplay between the solvent, electronic coupling, and temperature conditions on both quantum coherence and electronic transfer. We consider Methanol, the polar solvent used in~{\cite{Insuasty2011385}},  and, for completeness, we also quantify the effects due to a non polar solvent, Toluene. In both cases, the chromophore-solvent coupling is found to be in the strong regime and the corresponding non trivial non-Markovian dynamics is treated  within a non-perturbative open quantum system approach, with the numerically exact  HEOM method~{\cite{chen2015}}, on a spin-boson model of the complex-solvent~{\cite{RevModPhys,eckel2009,thorwart2009}}.

This paper is organised as follows: the chromophore and solvent  properties are developed in Sec.~\ref{sec:chromo_solvent}; the chemical structure and main parameters of the organic molecule are presented in Sec.~\ref{sub:chemical_structure}, and the mathematical model and chromophore+solvent physical parameters are given in Sec.~\ref{sub:hamiltonian}. The main results and discussions are presented in Sec.~\ref{sec:result}: solvent and temperature effects on the quantum dynamics (Sec.~\ref{sub:sol_temp_effect}), and electronic transfer (Sec.~\ref{sub:ETR}). Finally, conclusions are given in Sec.~\ref{sec:conclusion}.
\medskip

\section{Chromophore$+$Solvent System: Chemical and Physical Properties}\label{sec:chromo_solvent}

\begin{figure}[h]
\centering  
\includegraphics[width=\columnwidth]{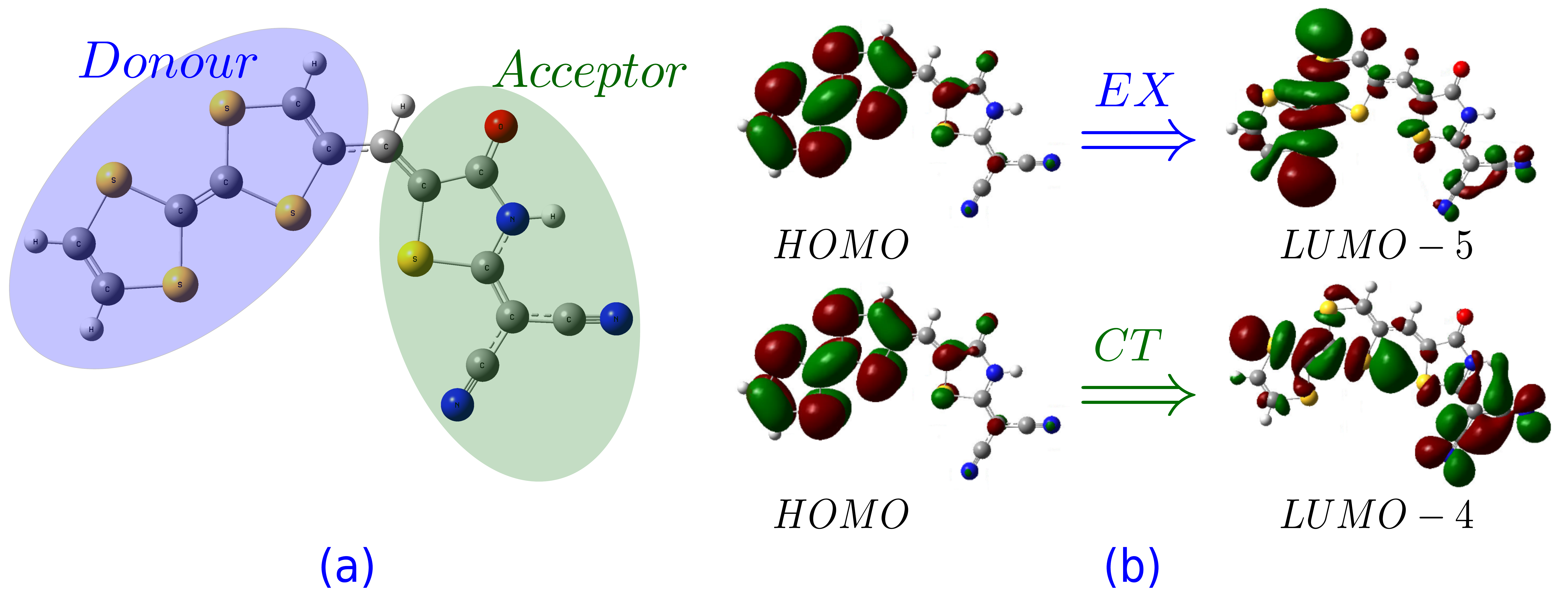}
\caption{(a) Molecular structure for the push-pull chromophore derived from Rhodanine; Dicyanorhodanine (Acceptor) +2-Formyltetratiafulvalene (Donor). (b) Electronic density associated to the EX and CT states involved in the ET process between D and A.}
\label{fig1}
\end{figure}

The considered push-pull chromophore derived from Rhodanine, the Dicyanorhodanine+2-formyltetratiafulvalene (D2F) complex~{\cite{Insuasty2011385,madrid2018}}, has a  chemical structure as shown in Fig.~{\ref{fig1}(a)}: it consists of a 2-formyltetratiafulvalene as the electron-donor (D) and a Dicyanorhodanine as the electron-acceptor (A) linked by Propylene. The compound exhibits appealing photo- and electron-chemical properties; a D$-$A intramolecular charge transfer in the absorption band of the donor at $\lambda_{max}=498$ nm (visible region). This was confirmed by solvatochromics studies using Acetone, Ethanol, and Methanol as solvents {\cite{Insuasty2011385}}: the ``push-pull''  compound displays a clear electrochemically amphoteric behaviour and  reveals a significant D$-$A electronic communication through the $\pi$-conjugated core, a feature that can be useful for ET integration into molecular systems of greater complexity.
\subsection{D2F Chemical Structure} \label{sub:chemical_structure}
The geometry of the D2F molecule was optimised by means of Gaussian 09~{\cite{gaussian09}}. The hybrid functional B3LYP and the 6-31G$^{++}$ basis are used to estimate the energy values of D and A electronic states, as well as the coupling between them. We consider a scenario where the system is firstly photoexcited from the ground electronic state to a $\pi$-$\pi^{*}$ state localised on D. This is followed by a nonradiative process that corresponds to an electronic transition from the $\pi$-$\pi^{*}$ state to a charge-transfer (CT) state that involves a significant D$-$A charge transfer. The molecular orbital analysis of both ground and excited states of the optimised geometry indicates that the HOMO (LUMO) is mainly located on D (A). The two aforementioned states are illustrated in Fig.~\ref{fig1}(b); the first transition that gives the excited state (EX) corresponds to the photoexcitation D--A $\rightarrow$ D$^{*}$--A at an excitation energy of $2.85$~eV. This exhibits an oscillator strength $f_{osc}=0.2348$ and a percentage contribution of 83\% (HOMO $\rightarrow$ LUMO+5 transition in Fig.~1(b)). The second one that generates the CT state (D$^{-}$--A$^{+}$) has an energy of $2.73$~eV, and this state has $f_{osc}=0.0002 $, with a percentage contribution of 75\% for the HOMO $\rightarrow$ LUMO+4 transition (Fig.~1(b)). These values were computed for Methanol via the time dependent density functional theory, by using the B3LYP functional. This scenario corresponds to an ET process from the 2-formyltetratiafulvalene compound to the Dicyanorhodanine  acceptor.

\begin{figure}[h]
\centering
\includegraphics[scale=0.13]{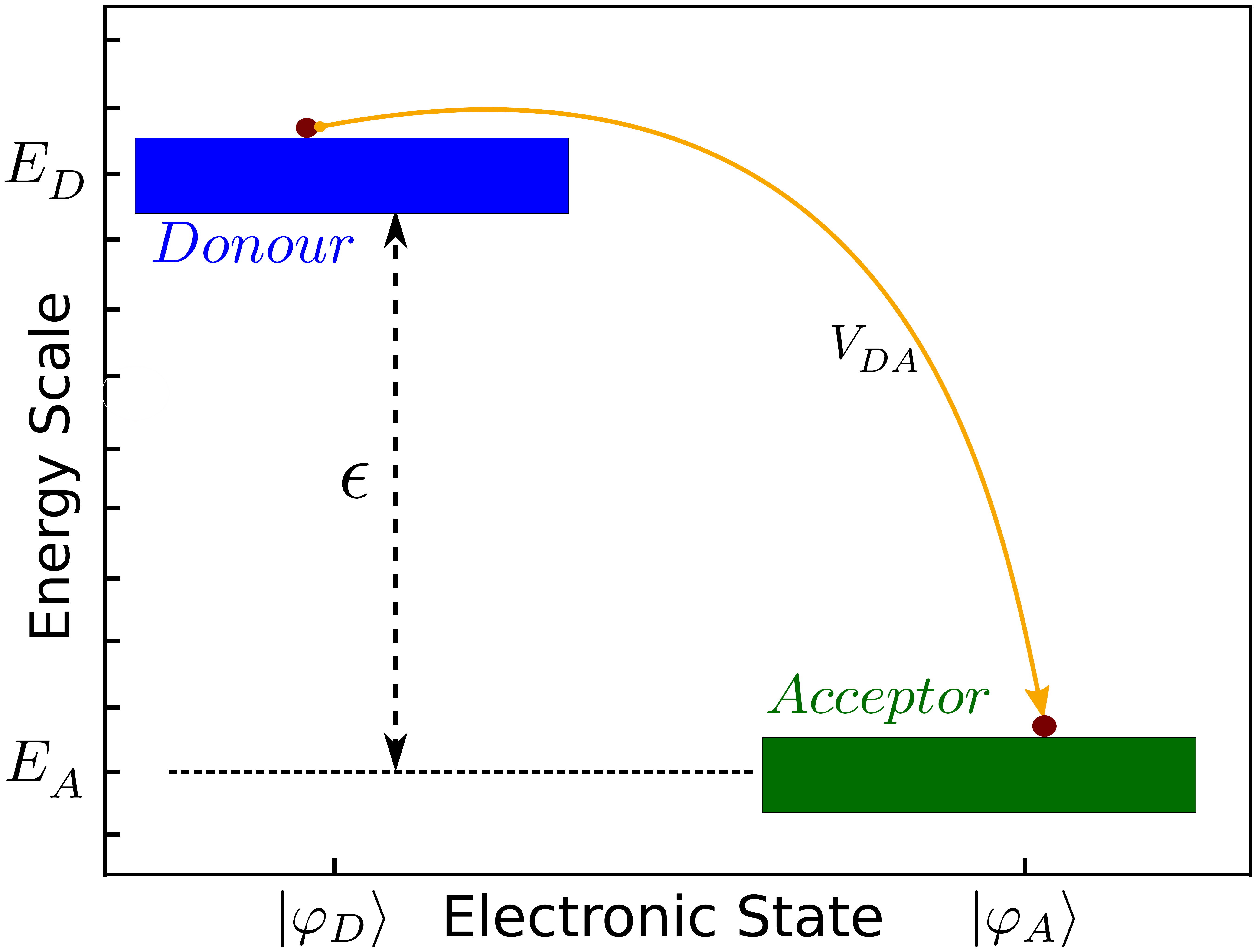}
\caption{Energy representation of the $\pi$-$\pi^{*}$ and CT states of interest in the D2F complex. $\epsilon\equiv |E_D-E_A|$ is the energy difference between the $EX$ and $CT$ states, and $V_{DA}$ stands for the D$-$A electronic coupling. Methanol:  $E_D=-4323$~cm$^{-1}$, $E_A=-5291$~cm$^{-1}$, $V_{DA}=280$~cm$^{-1}$; Toluene: $E_D=-5201$~cm$^{-1}$, $E_A=-5838$~cm$^{-1}$, $V_{DA}=317$~cm$^{-1}$.}
\label{fig3}
\end{figure}

The energies associated to the two main states involved in the ET mechanism are sketched in Fig.~\ref{fig3}. According to the B3LYP/6-31G$^{++}$ optimisation, the estimated site energies for Methanol are $E_D=-4323$~cm$^{-1}$ and $E_A=-5291$~cm$^{-1}$.  On the other hand, for Toluene $E_D=-5201$~cm$^{-1}$ and $E_A=-5838$~cm$^{-1}$.  Other physical parameters of interest are given in Fig.~\ref{fig3} and Table~\ref{tb:main_parameter}.
%
\begin{table}[h]
\caption{Chromophore$+$solvent physical parameters. Energy values are given by the B3LYP/6-31G$^{++}$ optimisation. Dielectric constants and dipole relaxation times are taken from {\cite{Horng, Shcherbakov, barthel1991}} for Methanol, and from {\cite{Cecilie2000, Vincent1967}} for Toluene. $\Delta\mu_M=1.1058\times10^{-29}$~Cm, $\Delta\mu_T=1.3455\times~10^{-28}$~Cm, $a_{0}=5.07\times 10^{-10}$~m, $\epsilon_{0}=8.8542\times10^{-12}$ ${\rm C^2/Nm^2}$. Subscripts $M$ and $T$ in the relative dipole moment, $\Delta\mu$, stand for Methanol and Toluene, respectively.}
\begin{center}
\scalebox{0.87}{
\begin{tabular}{|l|c|c|c|c|c|c|}
\hline
\multirow{2}{*}{Solvent}&\multicolumn{6}{c|}{Parameter}\\
\cline{2-7}
         & $\epsilon$ (cm$^{-1}$)  & $V_{DA}$ (cm$^{-1}$)&$\varepsilon_s$&$\varepsilon_{\infty}$&$\tau_D$ (ps)&$\tau_s$ (ps)\\
\hline
Methanol & $968$ & $280$ & $32.7$ & $1.8$ & $49.6$ & $0.1$ \\
\hline
Toluene  & $637$ & $317$ & $2.4$  & $2.2$ & $5.1$ & $0.2$ \\
\hline
\end{tabular}}
\end{center}   
\label{tb:main_parameter}
\end{table}

\begin{figure*}[ht]
\centering  
\includegraphics[scale=1.1]{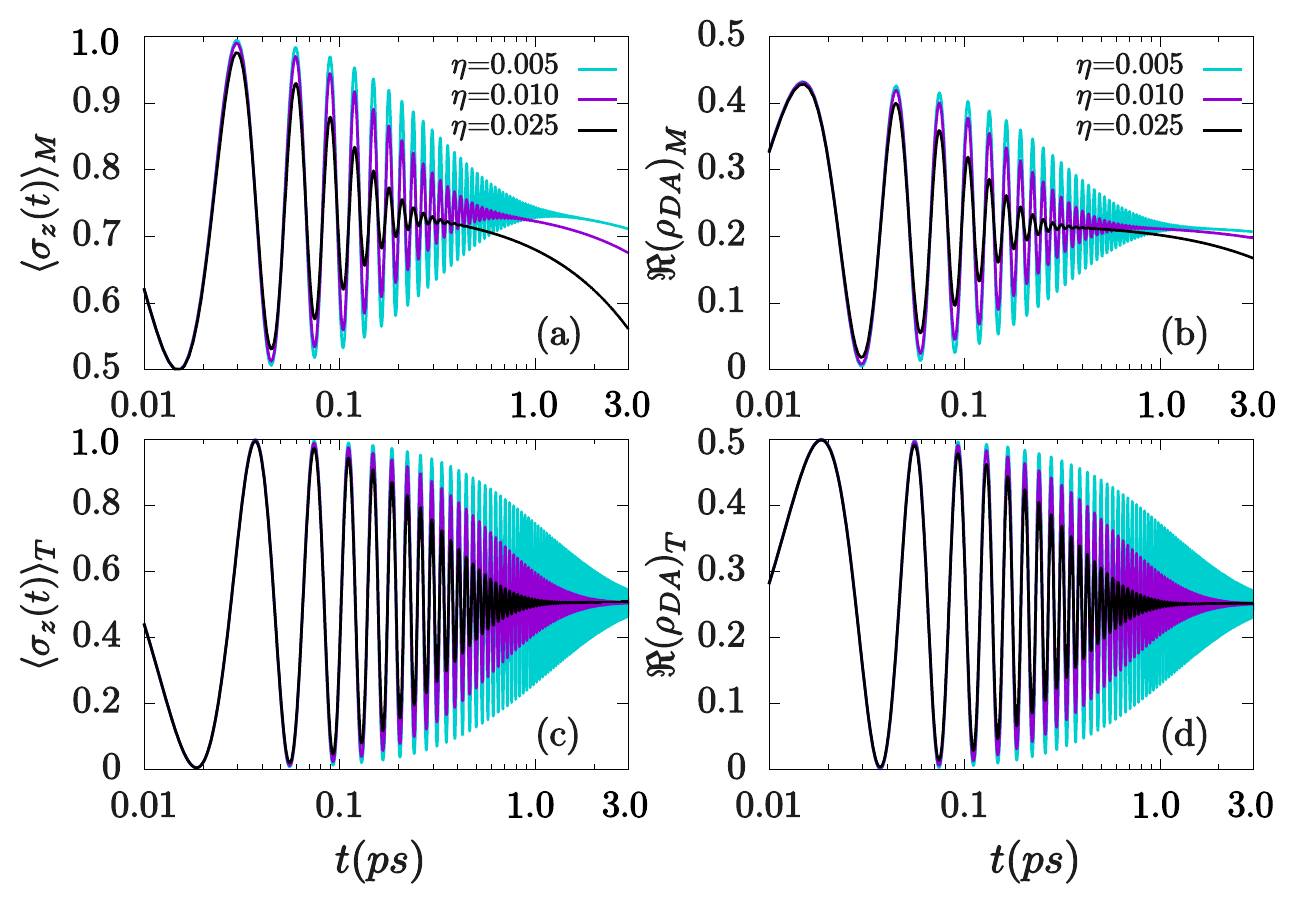}
\caption{\textit{Weak coupling regime}. Population inversion $\langle\sigma_z(t)\rangle$ (panels (a) and (c)) and real-part coherence $\Re{(\rho_{DA})}$ (panels (b) and (d)) dynamics as a function of the chromophore-solvent coupling at $T=300$~K. Subscripts $M$ and $T$ stand for Methanol and Toluene (top and bottom plots). In each figure, turquoise, violet, and black curves correspond to $\eta=0.005$, $\eta=0.010$, $\eta=0.025$, respectively. Hereafter, populations are considered to be localised on the donor at $t=0$, i.e., $\langle\sigma_z(0)\rangle=1$, and a logarithmic scale for the time axis is used in Figs. 3, 4 and 5.}
\label{fig4}
\end{figure*}

\begin{figure*}[ht]
\centering
\includegraphics[scale=1.1]{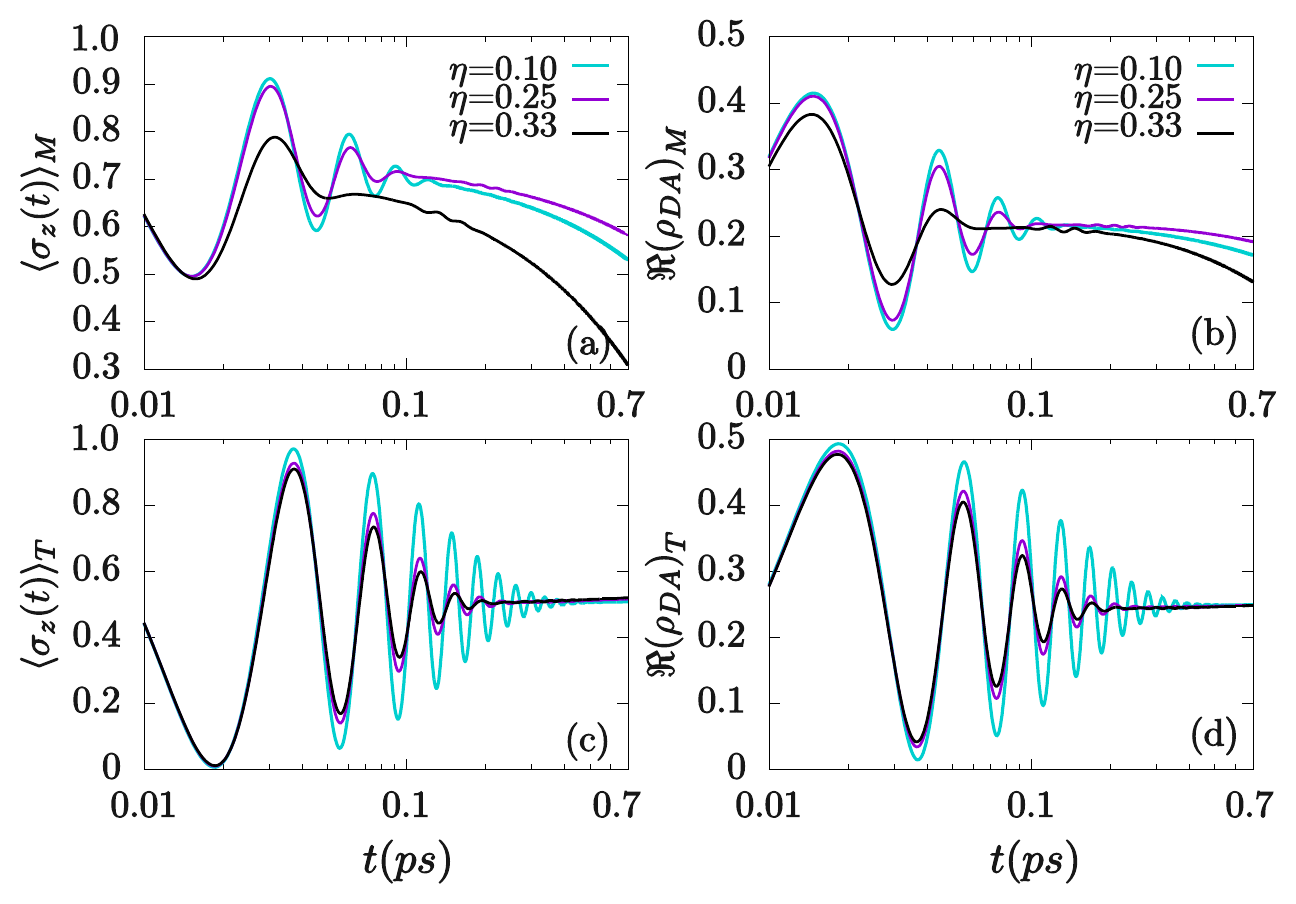}
\caption{\textit{Strong coupling regime}. Population inversion $\langle\sigma_z(t)\rangle$ (panels (a) and (c)) and real-part coherence $\Re{(\rho_{DA})}$ (panels (b) and (d)) dynamics as functions of the chromophore-solvent coupling at $T=300$~K. Methanol (top) and Toluene (bottom), for $\eta=0.10$ (turquoise), $\eta=0.25$ (violet), and $\eta=0.33$ (black) curves.}
\label{fig5}
\end{figure*}

\subsection{Hamiltonian and  Chromophore Quantum Dissipative Dynamics} \label{sub:hamiltonian}
We consider a spin-boson model to treat the chromophore-solvent ultrafast dynamics in a quantum fashion {\cite{Brisker, Malka, RevModPhys, eckel2009, thorwart2009}}. The total Hamiltonian can be written as $\widehat{H} = \widehat{H}_{S} + \widehat{H}_{B} + \widehat{H}_{SB}$, where the subscripts stand for system (the chromophore), bath (the solvent), and for the system-bath interaction, respectively. The solvent is modelled as a bath of harmonic oscillators with $\widehat{H}_{B} = \sum_{j}\hbar\omega_{j}\left( \widehat{a}_j^{\dagger}\widehat{a}_j + \frac{1}{2} \right)$, where $\widehat{a}_{j}^{\dagger}$ ($\widehat{a}_{j}$) denotes the creation (annihilation) operator of the $j$-th bosonic mode with frequency $\omega_{j}$. 

The $N$-partite donor-bridge-acceptor chromophore Hamiltonian is described in terms of the local (diabatic) electronic states $\ket{\varphi_n}$ at site $n$:
\begin{equation}
\widehat{H}_{S} = \sum_{n=1}^{N}E_{n}\vert \varphi_{n} \rangle\langle \varphi_{n} \vert + \sum_{n\neq m =1}^{N}V_{nm} \vert \varphi_{n} \rangle\langle \varphi_{m} \vert,
\label{Ecu2}
\end{equation}
where $E_{n}$ corresponds to the energy of the $n$-th site and $V_{nm}$ is the coupling energy between sites $n$ and $m$. Here, we consider a reduced two-site model ($N=2$) between the $EX$ and $CT$ states with energies as shown in Fig.~{\ref{fig3}}, following the prominent D$-$A charge transfer due to the $\pi$-$\pi^{*}$ state described in Sec.~\ref{sub:chemical_structure}. Thus, an effective two-level (D$-$A) system Hamiltonian can be identified for $\widehat{H}_{\text{S}}$:
\begin{equation}
\widehat{H}_{\text{TLS}}=\frac{\epsilon}{2}\widehat{\sigma}_z+V_{DA}\widehat{\sigma}_x,
\end{equation}
where $V_{DA}=\epsilon \mu_{12}/[(\mu_1-\mu_2)^2+4\mu_{12}^2]^{1/2}$
is the electronic coupling, here computed by means of the generalised Mulliken-Hush model \cite{cave1996,zheng2005}, $\epsilon\equiv|E_D-E_A|$ is the energy difference between donor and acceptor orbitals,  and $\widehat{\sigma}_{x,z}$ are Pauli's operators. For Methanol, $\mu_{12}=6.6990$ D is the transition dipole moment connecting the two adiabatic states in the charge transition, and $\Delta\mu_{12}=~\mu_1-\mu_2=18.8447$~D is the difference in adiabatic state dipole moments.  A similar procedure gives the corresponding values for Toluene.

The solvent is modelled by considering that each mode interacts independently with the electronic sites. The explicit interaction in the simplified two-site scenario, where $g_{j}$ accounts for the coupling strength to the $j$-th mode, is written as:
\begin{equation}
H_{SB} = \hbar\sigma_z\otimes\sum_{j}g_{j}(a_{j}^{\dagger} + a_{j}).
\label{eq:sb_int}
\end{equation}
 The information about the solvent and its interaction with the chromophore is encoded in the spectral density $J(\omega)=\sum_{j}g_j^2\delta(\omega-\omega_j)$ and the bath correlation functions (see Appendix \ref{app:method} for details). A precise description of this information through the spectral density is usually not a straightforward task. Here, we consider Onsager's reaction model for describing the D2F-solvent coupling. Geometrically, the D2F chromophore is assumed to occupy a region of radius $a_{0}$ ($\sim$ the van de Waals size) surrounded by the solvent~{\cite{Onsager}}. Within this model, a spectral density with Lorentzian shape can be derived~{\cite{Gilmore}}:
\begin{equation}
  J(\omega)=\frac{(\Delta\mu)^{2}}{2\pi\varepsilon_{0}a_{0}^{3}}\frac{6(\varepsilon_{s}-\varepsilon_{\infty})}{(2\varepsilon_{s}+1)(2\varepsilon_{\infty}+1)}\frac{\omega \tau_{s}}{\omega^{2}\tau_{s}^{2}+1},
  \label{Ecu1}
\end{equation}
where $\Delta\mu=\mu_{e}-\mu_{g}$ is the relative dipole moment between the excited and ground state dipole moments of the D$-$A molecule. $\varepsilon_{0}$, $\varepsilon_{s}$ and $\varepsilon_{\infty}$ stand for vacuum, solvent static and high frequency dielectric constants, respectively. $\tau_s$ is the solvent relaxation time (cutoff frequency $\omega_{c}\equiv1/\tau_{s}$). The reorganisation energy $\lambda:=\frac{2}{\pi}\int\frac{J(\omega)}{\omega}d\omega$ is simply given by $\lambda=2\hbar\alpha\omega_c$, where $\alpha = \frac{(\Delta\mu)^{2}}{4\pi\epsilon_{0}a_{0}^{3}}\frac{6(\epsilon_{s}-\epsilon_{\infty})}{(2\epsilon_{s}+1)(2\epsilon_{\infty}+1)\omega_c}$ is the frequency-independent factor of Eq.~\eqref{Ecu1}.

The quantum dissipative dynamics of the chromophore can be described within the density operator formalism~\cite{Weiss,breuer2002,RevModPhys}. If $\widehat\rho_T(t)$ is the time-dependent density operator associated to the total (chromophore$+$solvent) system (considered to be isolated), the reduced density operator for the chromophore is given by $\widehat\rho_S(t)=\text{Tr}_B(\widehat\rho_T(t))$. Here, we employ the HEOM technique {\cite{Yoshitaka, tanaka2010}} to numerically compute the temporal evolution of $\widehat\rho_S(t)$, as explained in Appendix~\ref{app:method}. 

The coherent nature of the ET depends on several factors such as temperature and chromophore-solvent coupling. Here, we consider two different solvents; Methanol (polar) and Toluene (non polar) and present numerical results for their effects on the D2F dynamics and the electronic transfer. Important physical parameters that characterise the D2F system are given in Table~{\ref{tb:main_parameter}}.
\section{Results and Discussion} \label{sec:result}

The spectral density {\eqref{Ecu1}} can be written in a simpler way as $J(\omega)=2\alpha\omega/(\omega^2\tau_s^2+1)$, where the  $\alpha$ values are determined from the parameters in Table~{\ref{tb:main_parameter}}, for each solvent. In fact, this chromophore-solvent coupling gives reorganisation energies of $86.18$~cm$^{-1}$ and $2.20$~cm$^{-1}$ for Methanol and Toluene, respectively. In order to directly compare the effects of both solvents on the D2F dynamics, we introduce a dimensionless system-bath coupling:
\begin{equation}
\eta := \frac{\alpha}{\hbar}=\frac{\lambda}{2\hbar\omega_c}.
\label{Ecu6}
\end{equation}
A system-bath coupling falls within the weak regime if $\eta\ll1$; in the strong regime $\eta\sim1$~\cite{Weiss,breuer2002,RevModPhys}. We obtained, according to the solvent properties given in Table {\ref{tb:main_parameter}},  $\eta=0.25$ and $0.33$ for Methanol and Toluene, respectively: these indicate that the strong coupling regime dictates the chromophore dynamics and ET in both cases. 
\begin{figure*}[ht]
\centering
\includegraphics[scale=1.1]{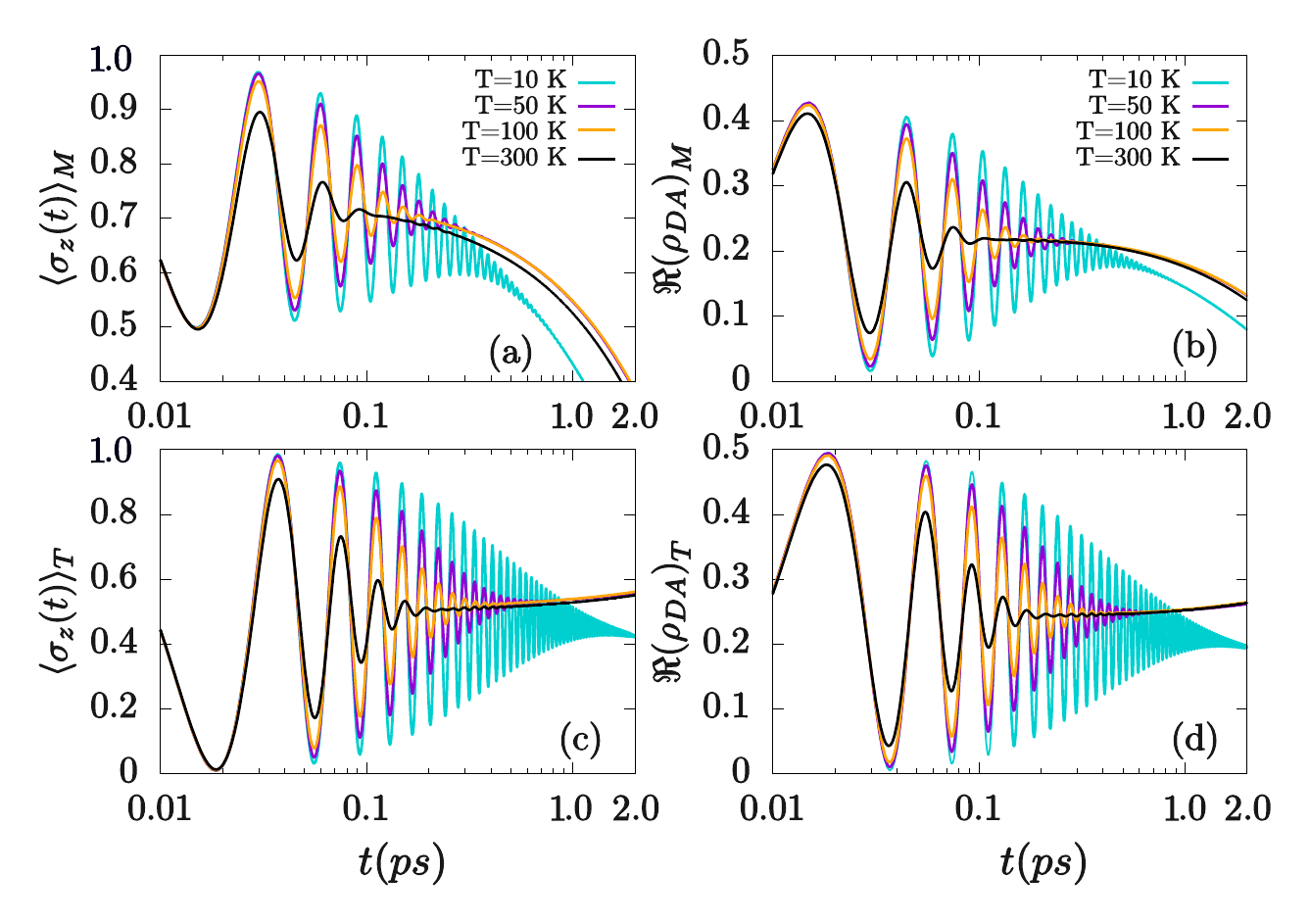}
\caption{\textit{Temperature dependence of D2F-solvent dynamics}. Population inversion $\langle\sigma_z(t)\rangle$ ((a) and (c)) and real-part coherence $\Re{(\rho_{DA}(t))}$ ((b) and (d)) dynamics as functions of temperature, for Methanol ($\eta_{M}=0.25$) and  Toluene ($\eta_{T}=0.33$). In each panel, turquoise, violet, orange and black curves correspond to $T=10$~K, $50$~K, $100$~K and $300$~K, respectively.}
\label{fig7}   
\end{figure*} 

\subsection{\bf Solvent and Temperature Effects} \label{sub:sol_temp_effect}

Having determined the main properties of the considered chromophore-solvent system, we investigate the solvent effects on the D2F dynamics in both weak and strong regimes. For doing so, we vary the value of $\eta$ and compute the population inversion $\langle\sigma_z(t)\rangle=\rho_D(t)-\rho_A(t)$, as well as the real-part coherence $\Re{(\rho_{DA}(t))}$ associated to the chromophore, where $\rho_D$, $\rho_A$ (populations) and $\rho_{DA}=\rho_{AD}^*$ (coherence) are the matrix elements of the density operator $\widehat\rho_S(t)$. 
In all the figures, populations are considered to be localised on the donor at $t=0$, i.e., $\langle\sigma_z(0)\rangle=1$, and a logarithmic time scale is used in Figs.~\ref{fig4}, \ref{fig5} and \ref{fig7}.

We begin by considering room temperature effects on the D2F dynamics, since possible applications of this chromophore as an efficient electron acceptor are expected to work at such condition. As shown in Fig.~{\ref{fig4}} (weak regime) and Fig.~{\ref{fig5}} (strong regime), these are computed for Methanol and Toluene  in terms of the chromophore-solvent coupling $\eta$. As expected, the oscillatory behaviour of population inversion and coherence persists for longer times in the weak regime. However, it is worth noting that those oscillations are more favoured in Toluene (than in Methanol), regardless the strength of the coupling regime. 

In the strong regime, say $\eta=0.25$ (see violet line in Fig.~{\ref{fig5}}) the population inversion and coherence oscillations disappear around $100$~fs in Methanol, while they continue to occur up to $\sim 220$~fs in Toluene. The process of electronic energy transfer is more coherent in Toluene as the interplay between the reorganisation energy and the inter-chromophore coupling lies in the coherent transfer regime $\lambda\ll V_{DA}${\cite{Ishizaki:2010}}. Indeed,  $\lambda/V_{DA}\sim0.7\times10^{-2}$ for Toluene while $\lambda/V_{DA}\sim0.3$ for Methanol (values of $\lambda$ are taken from Table~{\ref{tab2}}).

In spite of the fact that the dynamics in Methanol becomes incoherent faster than in Toluene, the final population inversion is more efficient in the former solvent; i.e., the stronger the coupling $\eta$ the higher the D-to-A transfer. In particular, the stationary population inversion in Toluene is $\sim25\%$, and independent of $\eta$ (panel {(c)} in Figs.~\ref{fig4} and {\ref{fig5}}). For Methanol, although the standing population inversion starts at a $\sim15\%$ in the weak regime, it increases with $\eta$ and reaches about a $\sim40\%$ in the strong regime (panel {(a)} of Figs.~{\ref{fig4}} and {\ref{fig5}}). This implies that a higher ET can be obtained in Methanol at room temperature. This is in agreement with the ET rate computed below in Section~\ref{sub:ETR} in terms of the density operator. We stress that this phenomenon takes place at room temperature and that the ET does not increase at lower temperatures, as shown below in Fig.~{\ref{fig8}}.

The studied D2F complex exhibits a long-term coherence that is comparable to the ones experimentally found  in some light-harvesting and organic materials~{\cite{lee2007C, collini2009, hwang2010, tempelaar2015, roscioli2017}}. As shown in Fig.~{\ref{fig5}(b)} for Methanol, coherence persists above $\sim100$~fs in the strong coupling regime $\eta_M=0.25$,  where the system is found according to the computed reorganisation energy. Furthermore, the coherence lifetime is longer for Toluene ($\gtrsim 220$~fs, see Fig.~{\ref{fig5}(d)}). As expected, the decoherence process is slower for both solvents in the weak coupling regime (Figs.~{\ref{fig4}(b) and \ref{fig4}(d)}). 
 
 In addition to the time scale introduced by the solvent relaxation ($\tau_s \approx 100$~fs; $\hbar\w_c=334$~cm$^{-1}$ for Methanol, and $\tau_s\approx 220$~fs; $\hbar\w_c=167$~cm$^{-1}$ for Toluene), other relevant physical scales of interest are set by thermal fluctuations, the reorganisation energy,  and the intrinsic electronic coupling. In particular, thermal fluctuations set by $k_BT/\hbar$ introduce a time scale $\tau_{T}= \w_{T}^{-1}\equiv \hbar/k_BT$ that affects the chromophore dynamics for $t > \tau_{T}$. On the other hand, quantum vacuum fluctuations contribute to the dephasing process in the regime $\tau_s <t <\tau_{T}$. Solvent polarity also plays  a crucial role as it affects the electronic energy levels of the states involved in the transfer process. In our particular case, Methanol exhibits a separation between the considered electronic states higher than Toluene ($\epsilon_M>\epsilon_T$). A full numerical comparison of the different time scales is given in Table~\ref{tab2} by means of the associated energy  ratios.
 \begin{table}[htp]
\begin{center}
\centering
\caption{Relevant energy scales for the D2F-solvent: ratio among thermal, decoherence, and inter-chromophore coupling energies. The involved parameters are as in Table {\ref{tb:main_parameter}}.}
\scalebox{0.86}{
\begin{tabular}{|c|c|c|c|c|c|c|c|c|} 
\hline  
Solvent & $T$(K) & $\frac{\hbar\omega_{T}}{\hbar\omega_{c}}$ & $\frac{\epsilon}{\hbar\omega_{c}}$ & $\frac{V_{DA}}{\hbar\omega_{c}}$ & $\frac{\hbar\omega_{T}}{\epsilon}$ & $\frac{\hbar\omega_{T}}{V_{DA}}$ & $\frac{\epsilon}{V_{DA}}$ \\[1.5mm]\hline 
\hline
\multirow{4}{23mm}{\textbf{Methanol} $\eta=0.25$ $\lambda=85.80$ cm$^{-1}$}&  &      &      &      &       &      &         \\
     & 300 & 0.62 & 2.90 & 0.84 & 0.22 & 0.75 &  3.46 \\ 
     & 50  & 0.10 & 2.90 & 0.84 & 0.04 & 0.12 &  3.46 \\  
     & 10  & 0.02 & 2.90 & 0.84 & 0.01 & 0.03 &  3.46 \\
\hline
\multirow{4}{23mm}{\textbf{Toluene} $\eta=0.33$ $\lambda=2.20$ cm$^{-1}$} & &      &     &      &       &      &        \\
     & 300 & 1.25 & 3.81 & 1.90 & 0.33 & 0.66 &  2.01 \\ 
     & 50  & 0.21 & 3.81 & 1.90 & 0.06 & 0.11 &  2.01 \\  
     & 10  & 0.04 & 3.81 & 1.90 & 0.02 & 0.02 &  2.01 \\
\hline
\end{tabular}}
\label{tab2}
\end{center}
\end{table}

\begin{figure*}[ht]
\centering
\includegraphics[scale=1.1]{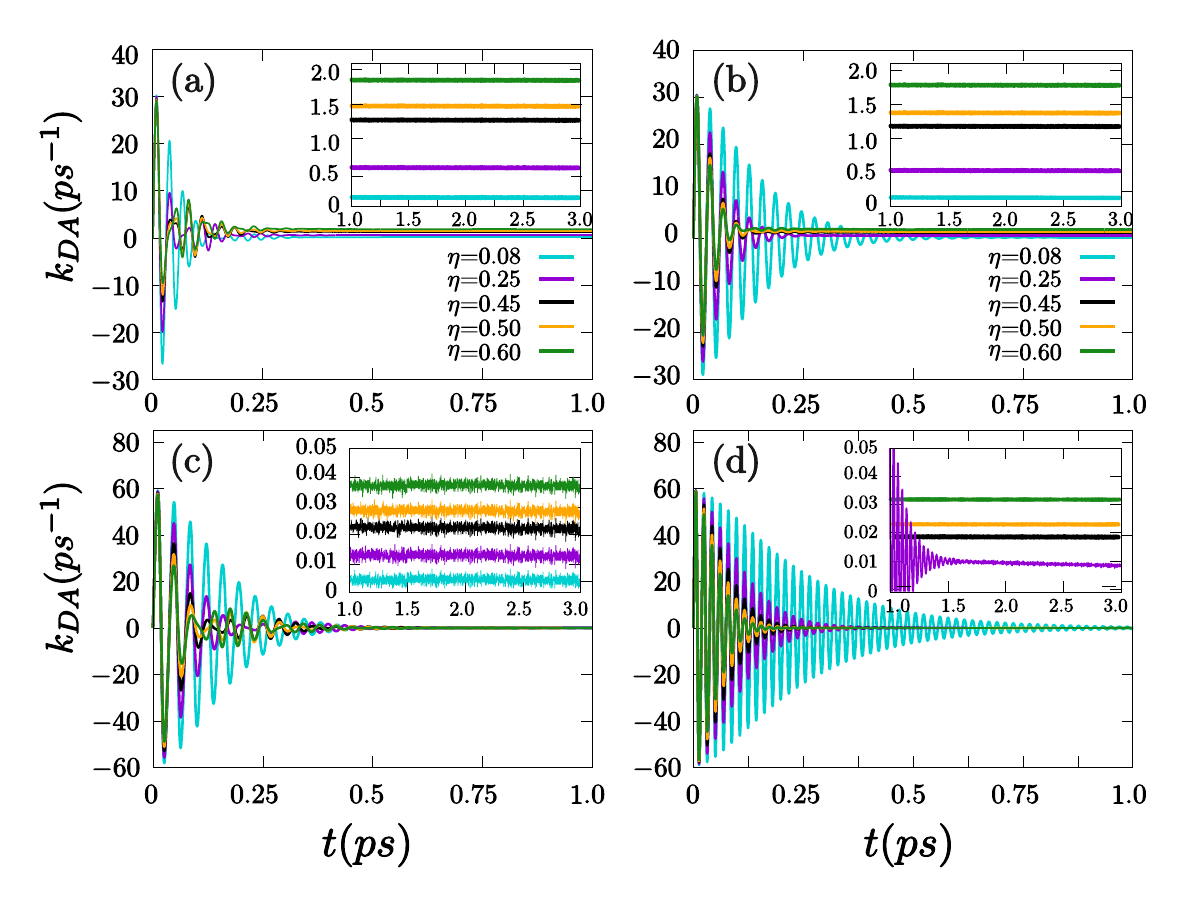}
\caption{\textit{Electron transfer rates as functions of the D2F-solvent coupling strength}. Methanol, $T=300$~K~(a), $T=50$~K~(b), and Toluene,  $T=300$~K~(c), $T=50$~K~(d). Turquoise, violet, black, orange and green curves correspond to $\eta=0.08$, $\eta=0.25$, $\eta=0.45$, $\eta=0.50$ and $\eta=0.60$, respectively. Other parameters are as in Table \ref{tb:main_parameter}.}
\label{fig8}
\end{figure*} 

Temperature effects on the population inversion and coherence for a fixed D2F-solvent strong coupling are shown in Fig.~{\ref{fig7}}. We consider the system-bath coupling that naturally identifies each solvent, i.e., $\eta_M=0.25$ for Methanol and $\eta_T=0.33$ for Toluene. In these Figures, the temperature is varied from $10$~K to $300$~K. $\langle\sigma_{z}(t)\rangle$ and $\Re{(\rho_{DA})}$ exhibit oscillations showing the quantum character of the D2F dynamics even at room temperature along tens of femtoseconds. Despite the shortness of this quantum effect, experimental techniques involving femtosecond pump-probe spectroscopy and quantum coherence control are in place to detect and control the reported ultrafast quantum dynamics~{\cite{lee2007C, collini2009, hwang2010, tempelaar2015, roscioli2017}}.

\subsection{Electron Transfer in the D2F Complex} \label{sub:ETR}

Electron transfer in the chromophore can be analysed throughout the chemical reaction rate, which is defined in terms of the flux-flux correlation function between the population of reactant and product states~{\cite{QUA56}}. Thus, it is possible to relate the ET speed with the time-dependent density matrix elements previously computed in Section~\ref{sub:sol_temp_effect}. In general, for an $N-2$-bridge system, the charge transfer from site $m$ to site $n$ can be described, within a charge hopping model {\cite{blumberger2015}}, by a jump rate $k_{mn} \equiv k_{m \rightarrow n}$, and the occupation probability (population) for each site can be written in terms of the ET rates by means of the following set of coupled differential equations:
\begin{eqnarray}
 &&\frac{d\rho_{D}}{dt}  = -k_{D2}\rho_D + k_{2D}\rho_2    \label{eq:multi_rates}  \\ 
&&\vdots  \nonumber \\ 
   &&\frac{d\rho_{N-1}}{dt} = k_{N-2N-1}\rho_{N-2}  + k_{AN-1}\rho_A \nonumber\\ 
   &&\;\;\;\;\;\;\;\;\;\;\;\,\,- (k_{N-1N-2}+k_{N-1A})\rho_{N-1}   \nonumber   \\ 
   &&\frac{d\rho_{A}}{dt} = k_{N-1A}\rho_{N-1} - k_{AN-1}\rho_{A} .         \nonumber
\end{eqnarray}

In our $N=2$-site model, this set reduces to: 
\begin{equation}
\frac{d\rho_{D}}{dt}  = -k_{DA}\rho_D + k_{AD}\rho_A, \label{eq:donor_rate}
\end{equation}
for the donor, and the corresponding equation for the acceptor is given by the constrain $\rho_D+\rho_A=1$. The behaviour of the ET rate is quantified by considering the deviation from equilibrium of the density operator, which for the donor reads $\Delta\rho_D(t)=\rho_D(t)-\rho_D(t\rightarrow\infty)$. It is known that a time-dependent ET rate can be obtained for the expected exponential decay of the deviation approaching to zero. By taking the time derivative to $\Delta\rho_D(t)$, applying Eq.~\eqref{eq:donor_rate}, and assuming $k_{AD}=0$ for the sake of simplicity, according to the D2F chromophore site energies, the ET rate can be written as:
\begin{equation}
k_{DA}=-\frac{1}{\rho_D(t)}\frac{d\rho_{D}(t)}{dt}.
\label{Ecu9}
\end{equation}
Although the value for the ET rate comes from taking the limit to infinity in Eq.~\eqref{Ecu9}, this equation allows us to interpret the ET speed in terms of the quantum-mechanical dynamics of the chromophoric system. In so doing, we plot  the time dependence of $k_{DA}$ as a function of the chromophore-solvent coupling strength in Fig.~\ref{fig8}. An oscillatory behaviour is observed for $k_{DA}$ before the ET rate reaches its stationary value. We point out that beyond the hopping model here described, the asymptotic $k_{DA}$ relies on the physical properties of the system: the D$-$A coupling, the reorganization energy, etc. Indeed, we have compared our findings for the charge hopping model with the results found by means of the flickering resonance model, and corroborated that these converge to the same result for the ET rate (for a discussion on both methods see e.g.~\cite{blumberger2015,zhang2014}). Thus, the matching with the computed values in the insets of Fig.~\ref{fig8} is confirmed (numerics not shown).
As anticipated in Sec.~\ref{sub:sol_temp_effect}, the ET rate is favoured by a strong D2F-solvent coupling, for both solvents. In particular, for Methanol, the ET rate increases up to one order of magnitude as the strength of the D2F-solvent coupling increases, as shown in the insets of Figs.~\ref{fig8}(a) and (b). On the other hand, such a variation is less appreciable for Toluene, as can be seen in Figs.~\ref{fig8}(c) and (d), which is in agreement with the quantum dynamics reported in Figs.~\ref{fig4} and \ref{fig5}. 

The results obtained in Fig.~\ref{fig8} are a strong indication that non-Markovianity assists the electronic transfer in the considered organic molecules, and we stress that such a transfer is optimised  at room temperature. On the other hand, our calculations also show that cooling down the system (see Figs.~\ref{fig8}(b) and (d)), although enhances the chromophore quantum coherence, translates into a slight reduction of the coupling-assisted ET rate, as can directly be compared  between left ($T=300$~K) and right ($T=50$~K) panels, for both Methanol and Toluene. Thus, we conclude that ET rate enhancement in the chromophore is a direct consequence of the natural strong coupling between the complex and the solvent, and that this is not necessarily due to its quantum coherence properties (cfr. Figs.~\ref{fig7} and~\ref{fig8}).
 
Although a large number of molecules is required for direct applications in functional OPV devices, our results contribute to the understanding of fundamental ultrafast processes and reveal the importance of considering the quantum nature of the involved early-stage  electron transfer, whose dynamics and efficiency determine the function of the OPV material.   
Further investigations of solvent effects can take place by contemplating other relevant aspects, e.g.,  different spectral densities. In the case of many-body scenarios, techniques like machine learning~{\cite{hase2017}} could also be applied to further  extend the scope of our findings. 

\section{Conclusions}\label{sec:conclusion}

We have investigated the ultrafast quantum dissipative dynamics and electronic transfer processes in an electron acceptor push-pull Rhodanine derivative chromophore (D2F complex). In so doing, we have computed the effects due to the chromophore environment by considering polar (Methanol) and non polar (Toluene) solvents, under different temperature conditions. The D2F chemical structure (and the identification of suitable excited  and charge transfer states) allowed us to evaluate, within the Onsager model of solvation, the complex-solvent coupling strengths $\eta$, which clearly set the dynamics within the strong coupling regime.  The corresponding non-Markovian features of the solvent were treated  within a non-perturbative open quantum system approach, with the numerically exact  HEOM method~{\cite{chen2015}}.

Our findings demonstrate the quantum character (coherent oscillations of the time-dependent density operator) of the D2F dynamics, even at room temperature, for both solvents. Particularly, that coherence is favoured in the case of Toluene thanks to the interplay between the reorganisation energy and the chromophore-chromophore coupling, in both complex-solvent coupling regimes. In contrast, Methanol exhibits a less coherent but higher ET rates, as it turns out that the stationary population inversion increases with coupling strength $\eta$. On the other hand, Toluene exhibits a steady population inversion that is invariant with respect to $\eta$.

To give a simple description of the ET process, we computed the ET rate $k_{DA}$ as the deviation from equilibrium of the density operator probabilities. As mentioned before, this  ET rate is significantly  enhanced by Methanol. Our simulations  show that, at room temperature, the steady value of $k_{DA}$ increases with the coupling strength $\eta$, and reaches values almost two orders of magnitude higher in Methanol than in Toluene. 
Furthermore, although prolonged coherence oscillations are observed at lower temperatures, the ET rate optimises at room temperature for both solvents as it reaches higher steady values (compare right and left panels in Fig.~\ref{fig8}).

Finally, we conclude that the non polar solvent enhances the quantum properties of the organic chromophore dynamics when compared to the polar solvent; yet, and perhaps counterintuitive, it is the latter that exhibits the higher ET rates. Even though the question as to 
 whether quantum coherence is crucial/helps electron transfer mechanisms in light harvesting chromophoric systems remains open to debate, our findings reveal not to be the case in the D2F-solvent complex  here considered.

\appendix

\section{HEOM for the reduced system dynamics}\label{app:method}

We briefly describe the Hierarchical Equations of Motion employed to obtain the dissipative dynamics of the reduced chromophoric system.
Considering the total system to be formed by the D2F molecule (reduced system, $S$) plus the solvent (bath, $B$), the time evolution of the associated density operator can be computed as~\cite{Weiss}: $i\hbar\dot{\widehat\rho}_{T}(t)=\mathcal{L}_{T}(t)\widehat{\rho}_T(t)\equiv \left[\widehat{H}_{T}, \widehat\rho_T(t) \right]$, where the total Hamiltonian $\widehat{H}_T\equiv \widehat{H}$ (Sec.~\ref{sub:hamiltonian}). Assuming an uncorrelated initial condition  $\widehat\rho_S(0)\otimes\widehat\rho_B$, with the bath in the thermal state $\rho_B=\exp(-\beta\widehat{H}_B)/Tr(\exp(-\beta\widehat{H}_B))$, $\beta=1/k_BT$ ($k_B$ is the Boltzmann constant), the reduced system dynamics is obtained by tracing out over the bath degrees of freedom:
\begin{equation}
\widehat{\rho}_S(t)= Tr_{B}\left(\widehat{U}(t)\widehat\rho_S(0)\otimes\widehat\rho_B\right),
\label{Ecu12}
\end{equation}
where \cite{Strumpfer1, May, strumpfer2011} $\widehat{U}(t)=\exp(-\frac{i}{\hbar}\int_{0}^{t}d\tau\mathcal{L}_{T}(\tau))$. Hence, bath effects are encoded in the correlation functions \cite{Strumpfer1,Okamoto}
\begin{equation}
C_{a}(t)=\langle u_{a}(t)u_{a}(0)\rangle_{B}=\frac{1}{\pi}\int_{0}^{\infty} d\omega J_a(\omega)\frac{e^{-i\omega t}}{1-e^{\beta\hbar\omega}},
\label{Ecu13}
\end{equation}
where $u_a(t)=\sum_{\xi}b_{a\xi}\chi_{\xi}(t)$ are the time-dependent bath coupling operators. They arise from the system-bath Hamiltonian \eqref{eq:sb_int} written in terms of the bath position coordinates $\chi_{\xi}$. In general, this can be written as $H_{SB}=\sum_{a,\xi}b_{a\xi}\vert \varphi_{a} \rangle\langle \varphi_{a}\vert \chi_{\xi}$. For the two-site model used in the main text, the coupling strength of Eq. \eqref{eq:sb_int} simply reads $g_{\xi}=b_{\xi}/\sqrt{2\hbar m_{\xi}\omega_{\xi}}$.

For a  Drude type 
spectral density (Eq.\eqref{Ecu1}), the correlation functions can be expanded as:
\begin{equation}
C_{a}^{\infty}(t)=\sum_{k=0}^{\infty}c_{ak}e^{-\nu_{ak}t},
\label{Ecu15}
\end{equation}
where $\nu_{a0} = \omega_{c}$ and $\nu_{ak}= \frac{2\pi k}{\beta\hbar}$ ($k\geq 1$) are the Matsubara frequencies \cite{ishizaki2005}, while the correlation coefficients \cite{Shi} 
$c_{a0}~=~\frac{\lambda_{a}\omega_{c}}{\hbar}\left[ \cot\left(\frac{\beta\hbar\omega_{c}}{2}\right) - i \right]$ and 
$c_{ak}=\frac{4\lambda_{a}\omega_{c}}{\beta\hbar^{2}}\left( \frac{\nu_{ak}}{\nu_{ak}^{2}-\omega_{c}^{2}} \right)$, for $k\geq 1$. The sum in Eq.  (\ref{Ecu15}) must be truncated at some finite level $K$; $\nu_{aK}e^{-\nu_{aK} t} \approx \delta(t)$.
Each exponential term in Eq.~(\ref{Ecu15}) introduces a set of auxiliary density operators $\widehat{\rho}_{\textbf{n}}$ that  are governed by the hierarchical structure of coupled equations of motion \cite{Strumpfer1, Ishizaki}:

\begin{widetext}
\begin{eqnarray}
\label{Ecu16} 
\frac{d\widehat{\rho}_{\textbf{n}}}{dt} &=& -\frac{i}{\hbar}\left[\widehat{H}_{S},\widehat{\rho}_{\textbf{n}}\right]-\sum_{a=1}^{N}\sum_{k=0}^{K}n_{ak}\nu_{ak}\widehat{\rho}_{\textbf{n}}-i\sum_{a=1}^{N}\sum_{k=0}^{K}\left[\vert \varphi_{n}\rangle\langle \varphi_{n}\vert,\widehat{\rho}_{\textbf{n}_{ak}^{+}}\right] \nonumber 
\\ &&
-\sum_{a=1}^{M}\left(\frac{2\lambda_{a}}{\beta\hbar^{2}\omega_{c}} -\sum_{k=0}^{K}\frac{c_{ak}}{\nu_{ak}}\right) \big[ \vert \varphi_{a} \rangle\langle \varphi_{a} \vert,\big[ \vert \varphi_{a} \rangle\langle \varphi_{a} \vert,\widehat{\rho}_{\textbf{n}} \big] \big] \nonumber   \\ &&
- \sum_{a=1}^{N}\sum_{k=0}^{K}n_{ak}\left(c_{a,k}\vert\varphi_{a}\rangle\langle \varphi_{a} \vert \widehat{\rho}_{\textbf{n}_{ak}^{-}} + \widehat{\rho}_{\textbf{n}_{ak}^{-}}\vert \varphi_{a} \rangle\langle \varphi_{a} \vert c_{a,k}^{*} \right),
\end{eqnarray}
\end{widetext}
where each auxiliary density operator with nonnegative subscript $\textbf{n}=(n_{10}, \dots, n_{1K}, \dots, n_{N0}, \dots, n_{NK})$ is coupled to the auxiliary density operators with subscripts $\textbf{n}_{ak}^{\pm}=(n_{10}, \dots, n_{ak}\pm1, \dots, n_{NK})$. Each density operator in the hierarchy is assigned to a hierarchy level $L=\sum_{a=1}^{N}\sum_{k=0}^{K} n_{ak}$. The density operator $\widehat\rho_S$ of the reduced system of interest corresponds to that one with subscript $\textbf{n}=(0, \dots, 0)$.

The amount of auxiliary density operators (which rapidly increases with level $L$ and in principle is infinite) accounts for the non-Markovian character of the system dynamics. Hence, a truncation such that $L_{\text{max}}\gg\omega_{\text{max}}/\min{(\nu_{ak})}$, with $\omega_{\text{max}}$ being a characteristic frequency of the system of interest, must be taken. Convergence of the HEOM is finally achieved after specifying the way to cutoff the terms beyond the truncation $L_{\text{max}}$. For doing so, the so-called time-local truncation can be used such that the Markovian approximation is employed for auxiliary density operators at level $L_{\text{max}}$. This implies that, for all auxiliary density operators with $L=L_{\text{max}}-1$,
\begin{equation}
\sum_{k=0}^{K}\widehat{\rho}_{\textbf{n}_{ak}}^{+}   \approx    -i\left( \widehat{Q}_{a}^{K}(t)\widehat\rho_{\textbf{n}}-\widehat\rho_{\textbf{n}}\widehat{Q}_{a}^{K}(t)^{\dagger} \right),
\end{equation}
where $\widehat{Q}_{a}^{K}(t)= \int_{0}^{t}C_a^K(\tau)e^{\left( -\frac{i}{\hbar}\widehat{H}_{S}\tau\right)}\ketbra{\varphi_a}{\varphi_a}e^{\left(\frac{i}{\hbar}\widehat{H}_{S}\tau\right)}d\tau$.
\medskip

\section*{Acknowledgements}

We acknowledge J. Arce and A. Ortiz for helpful discussions. This work was supported by the Colombian Science, Technology and Innovation Fund-General Royalties System (Fondo CTeI-Sistema General de Regal\'ias) under contract BPIN 2013000100007. C.E.S. thanks Universidad de C\'ordoba for partial support (Grant No. CA-097).

\bibliographystyle{ieeetr}
\bibliography{Bibliography}

\end{document}